# Deep Beamforming for Speech Enhancement and Speaker Localization with an Array Response-Aware Loss Function

Hsinyu Chang, Yicheng Hsu, and Mingsian R. Bai, *Senior Member, IEEE*

**Abstract**: Recent research advances in deep neural network (DNN)-based beamformers have shown great promise for speech enhancement under adverse acoustic conditions. Different network architectures and input features have been explored in estimating beamforming weights. In this paper, we propose a deep beamformer based on an efficient convolutional recurrent network (CRN) trained with a novel ARray RespOnse-aWare (ARROW) loss function. The ARROW loss exploits the array responses of the target and interferer by using the ground truth relative transfer functions (RTFs). The DNN-based beamforming system, trained with ARROW loss through supervised learning, is able to perform speech enhancement and speaker localization jointly. Experimental results have shown that the proposed deep beamformer, trained with the linearly weighted scale-invariant source-to-noise ratio (SI-SNR) and ARROW loss functions, achieves superior performance in speech enhancement and speaker localization compared to two baselines.

*Index Terms*—Multichannel speech enhancement, speaker localization, loss function, deep learning

## I. INTRODUCTION

SPEECH enhancement (SE) aims at extracting the clean speech signals from the noisy mixture, which is essential for various applications such as hands-free communication, hearing aids, teleconferencing, etc. However, under adverse acoustic conditions such as reverberation and interference, the enhancement performance can be significantly degraded. Thanks to the advent of deep neural network (DNN) technology, learning-based monaural SE algorithms [2]–[6] have emerged with great promise in noise reduction.

DNN-based beamformers can be divided into two categories. One category is to integrate the DNN with a beamformer, referred to in this study as the two-stage weight estimation approach [7]–[12]. In the first stage, the spatial covariance matrices (SCM) of speech and noise signals are computed through time-frequency (T-F) masking estimated by a DNN. The computed SCMs are then used in the second stage to compute array weights according to various optimal beamforming design criteria [13]–[15]. However, numerical instability may arise if matrix inversion is required. To mitigate this problem, an All Deep Learning MVDR (ADL-MVDR) network is proposed in [16], where the matrix operations are replaced by two recurrent neural networks (RNNs). Another category [17] attempts to estimate array weights directly through the DNN. Many DNN architectures have been suggested for estimating optimal filter weights, e.g., the multiple-in-multiple-out (MIMO) U-net structure [18] and the complex-valued spatial autoencoder (COSPA) structure [19]. Several input features that carry spatio-spectral information for weight estimation have also been investigated [20]–[23]. However, these learning-based methods focus only on speech enhancement and do not consider localization issues.

Chen *et al*. [24] integrate an auxiliary localization module into MIMO-Deep Complex Convolution Recurrent network (MIMO-DCCRN) to perform speech enhancement and localization jointly. The signal processing-based localization module (SPLM) and the neural localization module (NLM) are compared under different conditions. However, both localization modules require grid search. A localization error may occur if the speaker is not located at one of the preselected grid points.

In this study, we propose a deep beamformer capable of jointly performing speech enhancement and speaker localization. The system is based on a convolutional recurrent network (CRN) [5]. Instead of using an auxiliary module NLM as in [24], we train the DNN with a loss function of weighted objectives including a scale-invariant source-to-noise ratio (SI-SNR) and an array response-aware (ARROW) loss. From the point of view of array signal

This work was supported by the National Science and Technology Council (NSTC), Taiwan, under the project number 110-2221-E-007-027-MY3.

Hsinyu Chang is with the Department of Electrical Engineering, National Tsing Hua University, Hsinchu, Taiwan (e-mail: sunny88081228908051@gmail.com)

Yicheng Hsu is with the Department of Power Mechanical Engineering, National Tsing Hua University, Hsinchu, Taiwan (e-mail: shane.ychsu@gmail.com).

Mingsian R. Bai is with the Department of Power Mechanical Engineering and Electrical Engineering, National Tsing Hua University, Hsinchu, Taiwan (e-mail: msbai@pme.nthu.edu.tw).

processing [25], the ARROW loss adopts the ground truth relative transfer functions (RTFs) of the target speaker and interferer for better enhancement and localization performance. In particular, the weighting parameters used in the ARROW loss function are thoroughly examined from the perspectives of enhancement and localization. The main contributions of this paper can be summarized as follows:
1) We present a combination of SI-SNR and ARROW loss functions designed for multichannel speech enhancement and speaker localization.
2) We investigate the impact of that the weighting parameters in the proposed loss function on speech enhancement and speaker localization.
3) We show that the introduction of the ground truth RTFs improves the performance and the robustness of localization in the presence of unseen room impulse responses (RIRs).

The remainder of this paper is organized as follows. In Sec. II, the problem formulation and the signal model are introduced. In Sec. III, the proposed method is presented in detail. The experimental setup and results are described in Secs. IV and V. The paper is concluded in Sec. V.

## II. PROBLEM FORMULATION AND SIGNAL MODEL

Consider an array of $M$ microphones receiving speech signal and noise signal from a farfield speaker and an interferer. The noisy signal $\mathbf{Y} \in \mathbb{C}^{M \times 1}$ captured by the microphone array can be written in the short-time Fourier transform (STFT) domain as

$$\mathbf{Y}(l,f) = \mathbf{R}_s(f)S(l,f) + \mathbf{R}_n(f)N(l,f) + \mathbf{v}(l,f) \quad (1)$$

where $S(l,f)$ and $N(l,f)$ denote the target speech signal and the interferer corresponding to the frequency bin index $f$ and the time frame index $l$, $\mathbf{R}_s \in \mathbb{C}^{M \times 1}$ and $\mathbf{R}_n \in \mathbb{C}^{M \times 1}$ denote the relative transfer functions (RTFs) associated with the target speaker and the interferer, respectively. $\mathbf{v} \in \mathbb{C}^{M \times 1}$ denotes the noise term comprising diffuse noise such as late reverberation.

We seek to enhance the signal $\hat{S}$ by using a filter-and-sum beamformer with array weights, $\mathbf{W} \in \mathbb{C}^{M \times 1}$:

$$\hat{S}(l,f) = \mathbf{W}^H(l,f)\mathbf{Y}(l,f) \quad (2)$$

where superscript "$H$" denotes the conjugate-transpose operator.

## III. PROPOSED SYSTEM

Figure 1 shows the DB system diagram, where a DNN is used to directly estimate the beamforming weights for subsequent enhancement and localization. In the training phase (indicated by the dashed blue box), the ground truth RTFs, the time-frequency domain target speaker signal, and the time-domain target speech received by the reference microphone are used to compute the weighted loss, as detailed next.

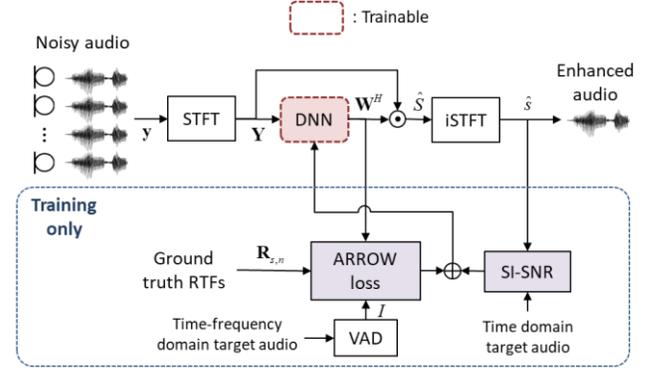

Fig. 1. The proposed deep beamformer. $\odot$ and $\oplus$ indicate the operations of Eq. (2) and Eq. (7). The processing units in the dashed blue box are used only in the training phase.

### A. Loss Function

To perform jointly learning-based enhancement and localization, we propose an ARray RespOnse-aWare (ARROW) loss function for training the DNN unit in Fig. 1. We motivate the development of the ARROW by starting with the scale-invariant source-to-noise ratio (SI-SNR) [26] loss function for the multichannel speech enhancement:

$$\mathcal{L}_{\text{SI-SNR}} = -10\log_{10}\frac{\|\eta \mathbf{s}\|_2^2}{\|\hat{\mathbf{s}} - \eta \mathbf{s}\|_2^2}, \quad \eta = \frac{\langle \hat{\mathbf{s}}, \mathbf{s} \rangle}{\|\mathbf{s}\|_2^2} \quad (3)$$

where $\{\hat{\mathbf{s}}, \mathbf{s}\} \in \mathbb{R}^{1 \times T}$ are the vectors of the inverse STFTs of $\{\hat{S}(l,f), S(l,f)\}$, respectively, $\langle \cdot \rangle$ denotes the inner product between two vectors, and $\|\cdot\|_2$ is the Euclidean norm. Using the array signal model in Eq. (1), the equivalent optimal solution of the SI-SNR in the frequency domain can be written as

$$\begin{aligned}\hat{S}(l,f) &= \mathbf{W}^H(l,f)\mathbf{Y}(l,f) = \eta S(l,f) \\ \Rightarrow &\left[\mathbf{W}^H(l,f)\mathbf{R}_s(f) - \eta\right]S(l,f) \\ &+ \mathbf{W}^H(l,f)\left[\mathbf{R}_n(f)N(l,f) + \mathbf{v}(l,f)\right] = 0\end{aligned} \quad (4)$$

Thus, minimizing the SI-SNR loss would only partially fulfill the distortionless constraint

$$\mathbf{W}^H(l,f)\mathbf{R}_s(f) \approx \eta \quad (5)$$

with some of the effort going into reducing the interference and noise.

To further improve the enhancement performance and to provide localization information, an ARray RespOnse-aWare (ARROW) loss function is introduced as follows:

$$\mathcal{L}_{\text{ARROW-}\alpha} := \alpha \frac{1}{L_{tp}F} \sum_{lf} \left[ I(l) \left| \text{Im}\{\mathbf{W}^H(l,f)\mathbf{R}_s(f)\} \right| \right] +$$
$$(1-\alpha)\frac{1}{L_{ta}F}\sum_{lf}\left[(1-I(l))\left(\left|\text{Re}\{\mathbf{W}^H(l,f)\mathbf{R}_n(f)\}\right|\right.\right. \tag{6}$$
$$\left.\left. + \left|\text{Im}\{\mathbf{W}^H(l,f)\mathbf{R}_n(f)\}\right|\right)\right],$$

where $\text{Re}\{\cdot\}$ and $\text{Im}\{\cdot\}$ denote the real and imaginary part operators, $I(l) \in \{0,1\}$ is the indicator of a voice activity detector (VAD), $\alpha \in [0,1]$ is a weighting factor that weights the target and interference terms, $L_{tp}$ and $L_{ta}$ are the number of frames corresponding to the target speech present and absent periods, and $F$ is the number of frequency bins. Note that the first term of the loss function in equation (6) is intended to "clean up" the imaginary part of the distortionless constraint in equation (5), while the second term is intended to further reduce the array response associated with the unwanted directional interference. A natural question is why the distortionless constraint is not directly incorporated into the loss function in Eq. (6). We found it difficult to train our DNN model with this setting due to the scaling problem and some potential conflicts with the SI-SNR loss.

To formulate the complete loss function, we combine the SI-SNR and ARROW loss functions with linear weighting
$$\mathcal{L} = \beta \mathcal{L}_{\text{SI-SNR}} + (1-\beta)\mathcal{L}_{\text{ARROW-}\alpha} \tag{7}$$
where the weighting factor $\beta \in [0,1]$.

*B. Localization*

For localization of the target speaker, the following beampattern function is defined:
$$P(\theta) = \frac{1}{L_{tp}F}\sum_{f}\sum_{l} I(l)\left|\mathbf{W}^H(l,f)\mathbf{a}_\theta(f)\right| \tag{8}$$
where $\mathbf{W}(l,f)$ is the array weights obtained from DNN, $\mathbf{a}_\theta$ denotes the free-field plane-wave steering vector at the angle $\theta$ which ranges from 30° to 150° in 15° increments and $L_{tp}$ are the number of frames corresponding to the speech present periods. Note that we only consider the time when the target speaker is active ($I(l) = 1$).

It follows that the direction of arrival (DOA) of the speaker can be obtained by finding the peak of the beampattern function:
$$\hat{\theta}_s = \arg\max_{\theta} P(\theta) \tag{10}$$

*C. Deep Beamforming Network (DBnet)*

The DNN unit in Fig. 1 is implemented in a convolutional recurrent neural network (CRNN) architecture illustrated in Fig. 2, hereafter referred to as the deep beamforming network (DBnet). The beamformer weights can be estimated directly from the microphone signals using DBnet. The stacked real and imaginary parts of the microphone signals are the input data to the encoder. The decoder layer produces the array weights as output. In Fig. 2, the DBnet structure consists of four symmetric convolutional and deconvolutional encoder and decoder layers with a 16-32-64-64 filter. To reduce computational complexity, the separable convolution [27] is chosen for each convolutional block. Each convolutional block is followed by a batch normalization and ReLU activation. Tanh activation is used at the last layer. The 1×1 pathway convolutions are used with add-skip connections [5], which allows for considerable parameter reduction with little performance degradation. The bottleneck consists of a grouped linear (GLinear) [6] layer. A single 256-unit GRU layer is used to capture the temporal information.

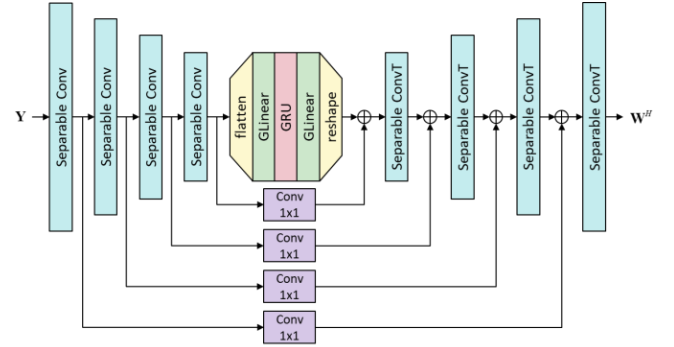

Fig. 2. The architecture of DBnet.

IV. EXPERIMENTAL STUDY

The proposed DB system is evaluated through the tasks of speech enhancement and speaker localization. To see the robustness of the proposed system to unseen acoustic conditions, we train our neural network using the simulated RIRs, but test it using the measured RIRs.

*A. Datasets*

Clean speech utterances are selected from the LibriSpeech corpus [28], where the subsets *train-other-500*, *dev-clean*, and *test-clean* are adopted for training, validation, and testing. The noise clip used as the directional interferer is selected from the Microsoft Scalable Noisy Speech Dataset (MS-SNSD) [29] and the Free Music Archive (FMA) [30]. In the MS-SNSD dataset, non-directional noise signal such as the babble noise is not included in the data preparation. Each training and testing signal mixture is prepared in the form of a 6-s clip randomly inserted with a 4-s clean speech clip. The training and validation sets comprise the signals with signal-to-interference ratio (SIR) randomly selected between -10 and 15 dB. The testing set consists of noisy signals with SIR = -5, 0, 5, and 10 dB. In addition, sensor noise is added with signal-to-noise ratio (SNR) = 20, 25, and 30 dB. A four-element uniform linear array (ULA) with an inter-element spacing of 8 cm is used in the experiment. Reverberant speech signals are simulated by convolving the clean signals with RIRs generated by the image source method [31]. Various reverberation times, (T60) = 0.2, 0.3, 0.4, 0.5, 0.6, and 0.7 s, are used. As illustrated in Fig. 3(a), the distance between the target speaker and interferer is randomly selected in the frontal plane at the ring sector bounded by radius = 0.75 and 2.1 m. In addition, any two sources are separated at least 15° apart from each other. The Multichannel Impulse Response Database [32], recorded at Bar-Ilan University using an eight-element ULA with an inter-element spacing of 8 cm for T60 = 0.16 s, 0.36 s, and 0.61 s, is adopted as the test set. In this study, we use only the RIRs of the four center microphones to generate the reverberant signals for testing. As shown in Fig. 3(b), the target speaker and the interferer appear randomly in any two of 9 angular directions equally spaced between 30° and 150° in 15° increments. A total of 30000, 3000 and 7200 samples are used for training, validation and testing.

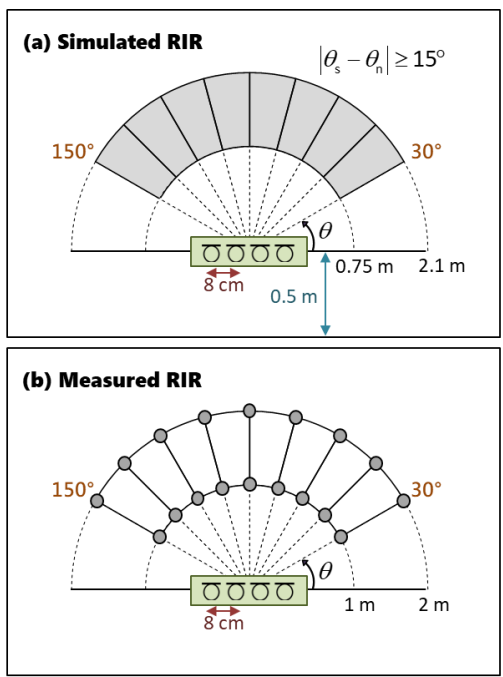

Fig. 3. Experimental setup for (a) training and (b) testing of the proposed deep beamformer.

### B. Baseline Methods

Two baselines are used for comparison with the proposed system. All models are implemented in the DBnet architecture. The first baseline is the DBnet trained with the SI-SNR loss. The second baseline is a DBnet cascaded with SPLM [24], trained with SI-SNR loss and binary cross-entropy loss as in [24]. This choice is made because SPLM does not require additional parameters. Here, SPLM-9 refers to the SPLM with 9 predefined zones. All datasets are generated at a sampling rate of 16 kHz. The signals are transformed to the STFT domain using a 25-ms Hamming window with a 10-ms hop size, and 512-point fast Fourier transform. The Adam optimizer is utilized in the training phase, with a learning rate of 0.001.

### C. Enhancement Performance

We use DNSMOS P.835 [33] to evaluate the speech enhancement performance. Three mean opinion scores based on P.835 human ratings are used to assess the speech quality (SIG), background noise quality (BAK), and overall quality (OVRL). First, we examine the effects of weighting $\beta$ between the SI-SNR loss and the ARROW loss on enhancement performance. As can be seen in Fig. 4, a large $\beta$ leads to an increased overall quality (OVL) and a signal quality (SIG) at the expense of increased background noise (BAK). Next, we examine the ARROW loss with different $\alpha$ factors, with a fixed weighting factor $\beta = 0.5$. The results in Fig. 5 show that the optimal enhancement performance is achieved when both weighting factors are set to 0.5. These results suggest that the target speech and the interference terms in the loss function are equally important for speech enhancement.

Next, we compare the enhancement performance of the proposed system when $\alpha = 0.5$, $\beta = 0.5$ with baselines. The results in Fig. 6 show that the proposed DB system performs the best in terms of all evaluation indices. Note that DBnet with SPLM performs worse than the original DBnet. This is due to the fact that the steering vector used in SPLM is based on the freefield plane wave model, which can lead to mismatch when applied to a reverberant environment. In summary, the method trained with the proposed ARROW loss can lead to much improved enhancement performance compared to the original DBnet method, by choosing appropriate weighting factors.

### D. Localization Performance

In this section, we evaluate the localization performance of the proposed DBnet with ARROW loss in comparison with two baselines (DBnet with SI-SNR loss and DBnet with SPLM).

To quantify the localization performance, we use the accuracy metric defined as

$$\text{Accuracy} = \frac{L_{true}}{L_{tp}} \times 100\% \tag{11}$$

where $L_{true}$ is the number of frames for which the angle estimation error is less than 15°, and $L_{tp}$ is the total number of frames with speaker active. As shown in Fig. 7, incorporating the ARROW loss results in superior speaker localization, with an average improvement of 5 %. In addition, the DBnet with the SPLM is outperformed by the DBnet trained with only SI-SNR loss due to the free-field steering vector used in training. Therefore, training the DBnet with the proposed ARROW loss allows for more robust localization than cascading with an SPLM.

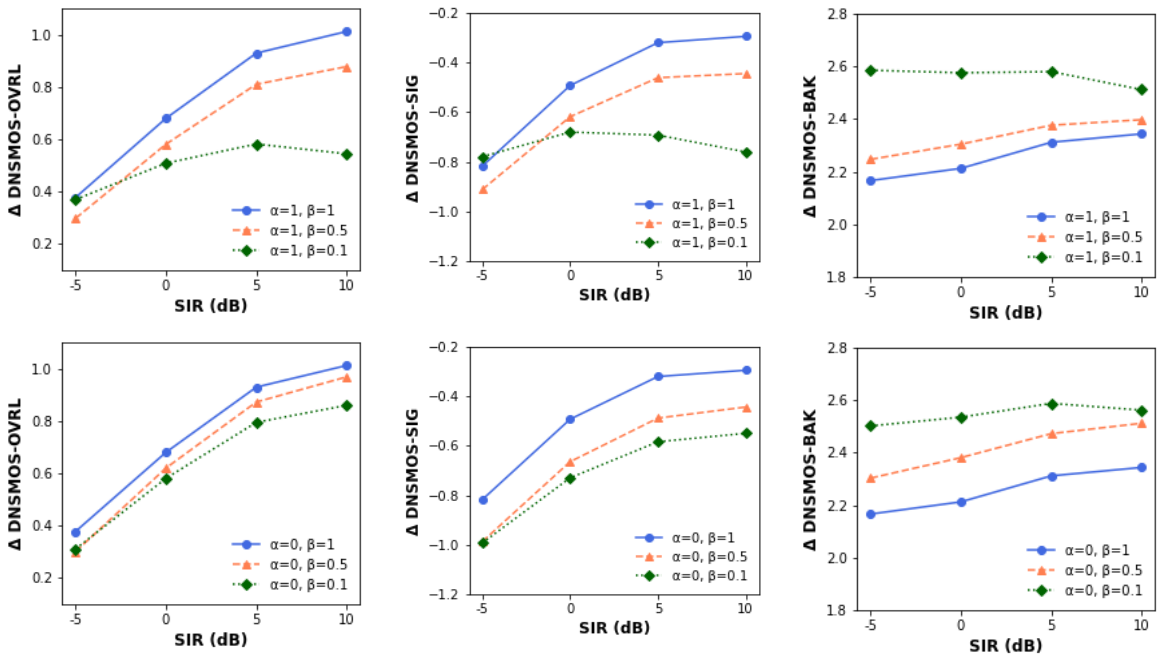

Fig. 4. Enhancement performance for different $\beta$ factors to weight SI-SNR and ARROW loss.

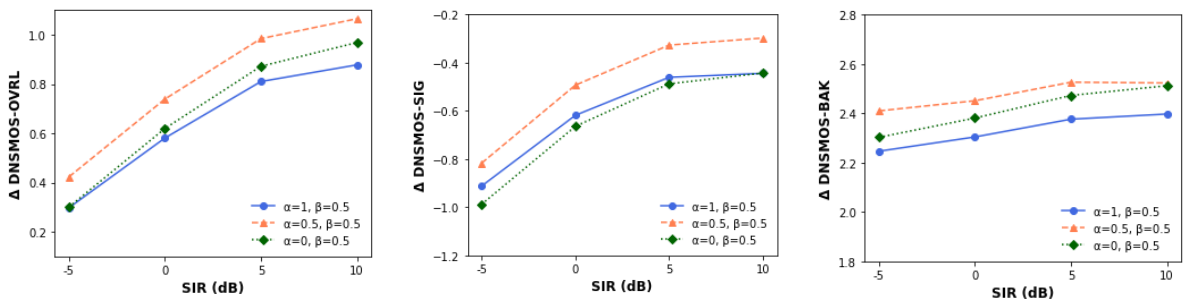

Fig. 5. Enhancement performance for different $\alpha$ factors to weight the ARROW loss.

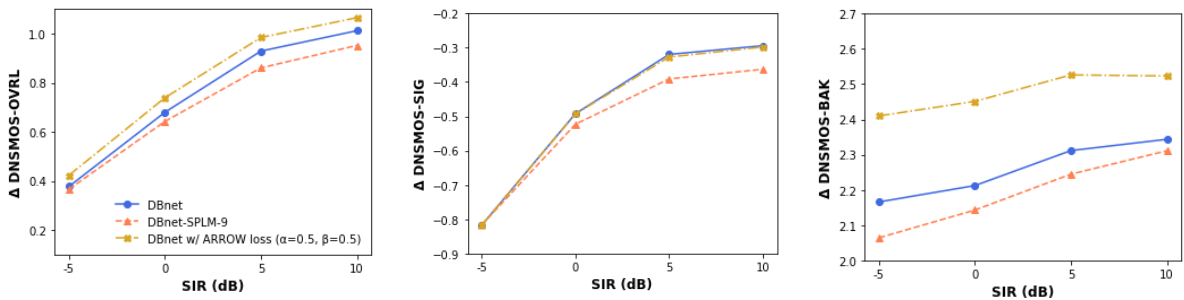

Fig. 6. Enhancement performance of the proposed method and the baselines.

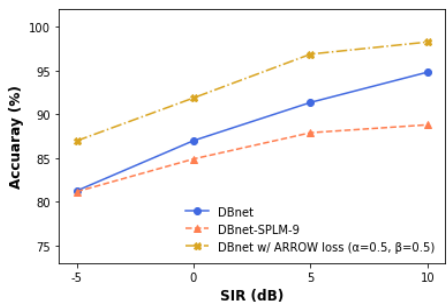

Fig. 7. Localization performance of the proposed method and the baselines.

I. CONCLUSIONS

In this study, we have proposed a deep beamforming system capable of speech enhancement and localization. A novel ARROW loss inspired by the distortionless constraint is proposed to effectively address these two tasks. The results have shown that the model trained with SI-SNR and ARROW loss provides superior enhancement and localization even when RIRs are not included in the training set. The future research agenda includes challenging scenarios with moving and multiple speakers.